\documentclass[showpacs,preprintnumbers,amsmath,amssymb,twocolumn]{revtex4-1}
\usepackage{graphicx}% Include figure files
\usepackage{dcolumn}% Align table columns on decimal point
\usepackage{bm}% bold math
\usepackage{subfigure}
\usepackage{float}
\usepackage{multirow}
\usepackage{booktabs}

\usepackage{amsmath}

\usepackage{latexsym,amssymb}
\usepackage[mathscr]{eucal}
\usepackage[switch*]{lineno}
\usepackage{cuted}
\usepackage[bookmarks=true,
   colorlinks=true,
   linkcolor=blue,
   urlcolor=blue,
   citecolor=blue,
   bookmarks=true,
   hyperindex=true
]{hyperref}

\begin{document}

\title{The new higher-order generalized uncertainty principle and primordial big bang nucleosynthesis}

\author{Song-Shan Luo\textsuperscript{1}}
\author{Zhong-Wen Feng\textsuperscript{1}}
\altaffiliation{Corresponding author: zwfengphy@cwnu.edu.cn}

\vskip 0.5cm
\affiliation{1 School of Physics and Astronomy, China West Normal University, Nanchong, 637009, China}

\begin{abstract}
As an important class of quantum gravity models, the generalized uncertainty principle (GUP) plays an important role in exploring the properties of cosmology and its related problems. In this paper, we explore the influence of the higher-order GUP on the primordial big bang nucleosynthesis (BBN). Firstly, based on a new higher-order GUP, we derived the Friedmann equations influenced by quantum gravity and the corresponding thermodynamic properties of the universe. Then, according to these modifications, we investigate BBN within the framework of GUP. Finally, combining the observational bounds of the primordial light element abundances, we constrain the bounds on deformation parameters of the new higher-order GUP. The results show that GUP has a significant effect on the BBN of the universe. Moreover, due to the unique properties of the higher-order GUP, it is found that value of the deformation parameter can be both positive and negative, which is different from the classical case.
\end{abstract}

\maketitle
% \linenumbers
\section{Introduction}
\label{sec1}
It is well known that general relativity is an important theory for describing the physical properties of macroscopic nature, while quantum theory plays an essential role in the study of physics at the microscopic scale \cite{Zeilinger,Will:2014kxa}. It has been hoped to merge these two theories into one, i.e., the construction of an effective theory of quantum gravity (QG). Therefore, gravitational quantization has always been a central issue in physics.

It is generally accepted in the study of QG that there exists a minimum measurable length with the Planck scale~\cite{Scardigli:1999jh,Maggiore:1993kv,Amelino-Camelia:2000stu}. Thus, combining the minimum measurable length with the classical physical  can lead to many new models of QG. For example, the combination of the minimum measurable length with Heisenberg's uncertainty principle makes the minimum measurable length effect a effective theory and plays a crucial role in gravitational physics. The  correction takes into account quantum effects at Planck scale, known as the generalized uncertainty principle (GUP)~\cite{Garay:1994en,Medved:2004yu,Amelino-Camelia:2004uiy,Khodadi:2017eim}. In recent years, the GUP has received significant attention  due to its applicability in physical systems with extremely small or high-energy scale~\cite{Chemissany:2011nq,Chen:2014jwq,Moradpour:2019wpj,Feng:2015jlj,Casadio:2020rsj,Iorio:2019wtn,Park:2020zom,Battista:2024gud}. The most obvious of these is the effect of GUP on the thermodynamics of black holes. In Refs.~\cite{Xiang:2009yq,Nozari:2012nf,Adler:2001vs}, the authors investigated the thermodynamic properties of different black holes in the framework of GUP. Their results showed that GUP can modify the Hawking temperature, specific heat, and entropy of black holes. Especially in the final stage of black hole evolution, the becomes very important because it may cause black holes to leave remnants of Planck size, meaning that Hawking radiation will not cause the black hole to completely evaporate, thus providing a solution to the black hole information paradox. Another important class of applications of GUP is the correction of the properties of cosmology. In Refs.~\cite{Zhu:2008cg,Feng:2022gdz,Anacleto:2015mma,Awad:2014bta,Salah:2016kre,Khodadi:2018wed,Khodadi:2018scn,Okcu:2020ybv}, it is found that the modified Friedman equations can be derived from the first law of thermodynamics and the GUP corrected entropy. Based on these works, one can easily analyze the impact of QG effects on the evolution of the universe and solve some important cosmological problems, such as baryon asymmetry and the inflation of the cosmology ~\cite{Luo:2023rhk,Das:2021nbq,Luciano:2021vkl,Okcu:2024tnw}.

On the other hand, the big bang nuclearsynthesis (BBN) is a theory that describes the sequence of nuclear reactions responsible for the synthesis of primordial light elements, such as $H$, ${^4}He$, $D$ and ${^7}Li$ \cite{Kolb:1981hk,Bernstein:1988ad}. The current accelerated expansion of the universe helps in the study of Big Bang cosmology. As we know, BBN occurs at a very early stage of the universe, and the extreme conditions at this time also allow for QG effects, so it is interesting to consider the GUP on BBN. In Ref.~\cite{Luciano:2021vkl}, according to the GUP model with a term
quadratic in the momentum $\Delta x\Delta p \geq {{\left[ {1 + 4\beta {{\left( {{{\Delta p} \mathord{\left/
 {\vphantom {{\Delta p} {{m_p}}}} \right.
 \kern-\nulldelimiterspace} {{m_p}}}} \right)}^2}} \right]\hbar } \mathord{\left/
 {\vphantom {{\left[ {1 + 4\beta {{\left( {{{\Delta p} \mathord{\left/
 {\vphantom {{\Delta p} {{m_p}}}} \right.
 \kern-\nulldelimiterspace} {{m_p}}}} \right)}^2}} \right]\hbar } 2}} \right.
 \kern-\nulldelimiterspace} 2}$ with the GUP parameter $\beta$ and the Planck mass $m_p$, Luciano obtained the modified Friedmann equations, and used them to  discuss the impact of QG effects on the BBN. Besides, he constrained the GUP parameter by using observational bounds from BBN and primordial abundances of the light elements. Note that the GUP model with a term quadratic in the momentum has some flaws. Such as the perturbations of these models are valid only for small values of the GUP parameter, and it  do not imply the non-commutative geometry. In order to resolve these limitations, the higher-order GUP models have been constructed. Considering that higher-order GUP have richer physical information, in this paper, we attempt to use a new higher-order GUP model with parameter adaptability for the minimum length to investigate the BBN. We first derive the modified Friedmann equations in the framework of new higher-order GUP. Next, according the GUP corrected Friedmann equations, the effect of GUP on BBN is analyzed. Especially, the GUP parameter $\beta_0$ is constrained by ensuring consistency between GUP cosmology predictions and the existing upper bounds on the variations of the freeze-out temperature $T_f$. Finally, we investigate GUP-induced deviations from standard cosmology on the primordial light element  abundances of $^{4} He$, $D$, and $^{4} Li$. Then, based on observational data of primordial abundances, we will constrain the GUP parameter $\beta_0$.

This paper is organized as follows. In section~\ref{sec2}, we  shortly review the new higher-order GUP and use it to derive the modified Friedmann equations. In section~\ref{sec3}, we explore the impact of GUP on BBN. In section~\ref{sec4}, combining the observations data of the primordial light element abundances, we constrain the bounds of deformation parameter $\beta_0$ of the new higher-order GUP. The work ends with conclusions and discussion in section~\ref{sec5}.

\section{High-order GUP and Friedmann equations}
\label{sec2}
In this section, we will briefly review the properties of the new higher-order GUP and its corresponding modified Friedmann equations. In Ref.~\cite{Du:2022mvr}, Du and
Long introduced a novel high-order generalized uncertainty principle model, which can be expressed as follows
\begin{align}
\label{eq1}
\Delta x\Delta p \geq \frac{\hbar }{2}\frac{1}{{1 \pm \left( {{{16\beta_0 \ell_p^2} \mathord{\left/ {\vphantom {{16\beta_0 \ell_p^2} {\Delta {x^2}}}} \right. \kern-\nulldelimiterspace} {\Delta {x^2}}}} \right)}}.
\end{align}
where $\beta_0$ is the GUP parameter, $\ell_p$ denotes the Planck length. In the following study, we will use units $\hbar  = c = {k_B} = 1$, which leads to the  ${\ell_p} = 1/{m_p} = \sqrt G$ with the gravitational constant $G$. Here, Planck mass $m_p \simeq 10^{19}$.  From Eq.~(\ref{eq1}), one can find this new high-order GUP has two main characteristics: one is the GUP parameter $\beta_0$ in the model can be assigned either positive or negative values. The symbol ``$+$" represents the positive GUP parameter, while ``$-$" represents the negative GUP parameter. The other one is that it maintains a fixed and uniform minimum length ${\rm{\Delta }}{x_{\text{min}}} = 4\sqrt {\left| \beta_0  \right|} {\ell_p}$ regardless of positive or negative parameters. These advantages ensure the role of QG effects in the model, allowing us to analyse the influence of positive or negative parameters on the same physical system.

On the other hand, in the holographic principle, when a gravitational system absorbs a particle, its area of apparent horizon and the total energy therein increase accordingly. Following the viewpoint in Refs.~\cite{Cai:2008ys,Bargueno:2015tea}, the minimal change of the area $\Delta A$ is given by
\begin{align}
\label{eq2}
\Delta A \sim X m,
\end{align}
where $X$ and $m$ denote the size and mass of the particle, respectively. Notably, in quantum mechanics, the standard deviation of the $X$ distribution is used to describe the width of the particle wave packet (i.e., $\Delta X$), which leads to a momentum uncertainty $\Delta p$ that cannot be larger than the mass. Therefore, Eq.~(\ref{eq2}) can be rewritten as $\Delta A \geq \Delta x \Delta p$, which implies that the minimum increment of the area of the gravitational system is limited by the momentum uncertainty $\Delta p$ and the position uncertainty $\Delta x$ of quantum mechanics. According Eq.~(\ref{eq1}), one can obtain the momentum uncertainty as follows
\begin{align}
\label{eq3}
\Delta p \geq \frac{\hbar }{2}\frac{1}{{\Delta x \pm \left( {{{16\beta_0 \ell _p^2} \mathord{\left/
 {\vphantom {{16{\beta_0} \ell _p^2} {\Delta x}}} \right. \kern-\nulldelimiterspace} {\Delta x}}} \right)}}.
\end{align}
For a static spherical gravitational system, the position uncertainty and the radius of the apparent horizon has the relationship ${\Delta x \approx 2r}$ \cite{Medved:2004yu}. Combining this relationship with the above equations, the minimal change of the area is $\Delta A \geq \chi \tilde \hbar \left( {{\beta_0}} \right)$, with the effective Planck constant $\tilde \hbar \left( {{\beta_0}} \right)$. It recovers  $\hbar  = {1 \mathord{\left/
 {\vphantom {1 2}} \right. \kern-\nulldelimiterspace} 2}$ for $\beta_0 = 0$.

Following information theory, the minimum increase of entropy is related to the value of the area, that is
\begin{align}
\label{eq4}
\frac{{{\text{d}}S}}{{{\text{d}}A}} \simeq \frac{{\Delta {S_{\min }}}}{{\Delta {A_{\min }}}} = \frac{1}{{8\tilde \hbar \left( {{\beta_0}} \right)}}.
\end{align}
In the classical model, the relationship between original entropy of a gravitational system and its horizon can be expressed as ${S_0} = {A \mathord{\left/ {\vphantom {A {4G}}} \right. \kern-\nulldelimiterspace} {4G}}$. If the area of horizon becomes a function of $A$, namely, $f\left( A \right)$, the entropy would change accordingly. Therefore, when considering the effects of higher-order GUP, the general expression of entropy becomes ${S} = {{f\left( A \right)} \mathord{\left/ {\vphantom {{f\left( A \right)} {4G}}} \right. \kern-\nulldelimiterspace} {4G}}$. Now, for obtaining the modified entropy, we need to employ derivative of this entropy with respect to area $A$ as follows
\begin{align}
\label{eq5}
\frac{{{\text{d}}S}}{{{\text{d}}A}} = \frac{{f'\left( A \right)}}{{4G}},
\end{align}
where $f'(A) = {{{\text{d}}f(A)} \mathord{\left/
 {\vphantom {{{\text{d}}f(A)} {{\text{d}}A}}} \right.
 \kern-\nulldelimiterspace} {{\text{d}}A}}$. Now, by comparing Eq.~(\ref{eq5}) with Eq.~(\ref{eq4}), one has
\begin{align}
\label{eq6}
f'(A)\left( {{\beta_0} > 0} \right) = \frac{1}{{2\tilde \hbar \left( {{\beta_0}} \right)}} = 1 + \frac{{16\pi {\beta_0}\ell _p^2}}{A},
\end{align}
and
\begin{align}
\label{eq7}
f'(A)\left( {{\beta_0} < 0} \right) = \frac{1}{{2\tilde \hbar \left( {{\beta_0}} \right)}} = 1 - \frac{{16\pi {\beta_0}\ell _p^2}}{A}.
\end{align}
By integrating Eq.~(\ref{eq6}) and Eq.~(\ref{eq7}), the modified entropy are ${S_{{\text{GUP}}}} = {A \mathord{\left/
 {\vphantom {A {4G}}} \right.
 \kern-\nulldelimiterspace} {4G}} + 4\pi \beta_0 \ell _p^2\ln \left( {{A \mathord{\left/
 {\vphantom {A G}} \right.
 \kern-\nulldelimiterspace} G}} \right)$ for $\beta_0 >0$, and ${S_{{\text{GUP}}}} = {A \mathord{\left/
 {\vphantom {A {4G}}} \right.
 \kern-\nulldelimiterspace} {4G}} - 4\pi \beta_0 \ell _p^2\ln \left( {{A \mathord{\left/
 {\vphantom {A G}} \right.
 \kern-\nulldelimiterspace} G}} \right)$ for $\beta_0 <0$, respectively.

 In Refs.~\cite{Feng:2022gdz,Luo:2023rhk,Cai:2008ys,Sheykhi:2010wm,Feng:2017vvw}, it is proved that the Friedman equations of FRW universe can be derived from the first law of thermodynamics and Bekenstein-Hawking entropy. Based on these works, considering a flat universe(the spatial curvature constant $k$=0), we have derived the Friedman equations in terms of modified entropy ${S_{{\text{GUP}}}}$ with the following expression
\begin{subequations}
\label{eq8}
\begin{align}
- 4\pi G\left( {\rho  + p} \right)  = \dot H f'\left( A \right),
 \\
\frac{8}{3}\pi G\rho   =  - 4\pi \int f' \left( A \right)\frac{{{\text{d}}A}}{{{A^2}}}.
\end{align}
\end{subequations}
where  $\dot H$, $\rho$ and  $p$ represent the derivative of the Hubble parameter $H$ with respect to time $t$, energy density and pressure of the universe, respectively. Substituting the expressions of $f'\left( A \right)$ into Eq.~(\ref{eq8}), and considering $A = {{4\pi } \mathord{\left/
 {\vphantom {{4\pi } {{H^2}}}} \right. \kern-\nulldelimiterspace} {{H^2}}}$, the modified Friedmann equations are given by
\begin{subequations}
\label{eq9}
\begin{align}
- 4\pi G\left( {\rho  + p} \right) & = \dot H\left( {1 \pm 4\beta_0 G{H^2}} \right),
 \\
\frac{8}{3}G\pi \rho & = {H^2} \pm 2G{H^4}\beta_0.
\end{align}
\end{subequations}
In above equations, the ``$+$" indicates that the modified Friedmann equations are affected the GUP of with $\beta_0 >0$, whereas ``$-$" represents the modified Friedmann equations are affected the GUP of with $\beta_0 < 0$. However, for $\beta_0 = 0$, the modifications return to the original cases. In the next section, we will employ the modified Friedmann equations~(\ref{eq9}) to investigate how the effects of QG influence the BBN.

\section{Big Bang Nucleosynthesis in the framework of GUP}
\label{sec3}
In this section, we attempt to investigate BBN in the framework of the higher-order GUP. According to the right hand side of modified Friedmann equation~(\ref{eq9}b), one has the the following relation
\begin{align}
\label{eq12}
H_{\beta_0} = {H}{\left( {1 \pm \frac{8}{3}{G^2}\pi \beta_0 \rho } \right)},
\end{align}
where $H_{\beta_0}$ and $H$ are the GUP corrected Hubble parameter and standard Hubble parameter, respectively. For the sake of simplicity, we can rewrite it in the following form~\cite{Luciano:2021vkl}
\begin{align}
\label{eq10}
H_{\beta_0}=H_0 Z_{\beta_0},
\end{align}
where ${H_0} = \sqrt {{{8G\pi \rho } \mathord{\left/
{\vphantom {{8G\pi \rho } 3}} \right. \kern-\nulldelimiterspace} 3}} $ is the classical Hubble parameter, and ${Z_{{\beta _0}}} = 1 \pm {{8{G^2}\pi {\beta _0}\rho } \mathord{\left/ {\vphantom {{8{G^2}\pi {\beta _0}\rho } 3}} \right. \kern-\nulldelimiterspace} 3}$ is the QG correction term for Hubble parameter. In BBN epoch, the universe is believed to be dominated by radiation. Therefore, the energy density of relativistic particles filling the universe satisfies $\rho  = {{{\pi ^2}g\left( T \right)T} \mathord{\left/ {\vphantom {{{\pi ^2}g\left( T \right)T} {30}}} \right. \kern-\nulldelimiterspace} {30}}$ with the degree of freedom of cosmic particles $g\left( T \right) \equiv {g_*} \sim 10$. In this case, $H_0$ and $Z_{\beta_0}$ in Eq.~(\ref{eq10}) can be rewritten as
\begin{align}
\label{eq15}
H_0 = \frac{{2{\pi ^{3/2}}\sqrt {{g_*}G} }}{{3\sqrt 5 }}{T^2},
\end{align}
\begin{align}
\label{eq16}
{Z_{\beta_0} } = 1 \pm \frac{4}{{45}}{g_*}{G^2}{\pi ^3}{T^4}\beta_0,
\end{align}
%\begin{align}
%\label{}
%{{\delta H}} = \frac{{2{\pi ^{3/2}}\sqrt {{g_*}G} {T}^2}}{{3\sqrt 5 }}\times\left( \pm {\frac{4}{{45}}{g_*}{G^2}{\pi ^3}{T}^4{\beta _0}} \right).
%\end{align}
For $\beta_0 >0$, the Eq.~(\ref{eq16}) takes the positive sign, while it takes the negative sign for $\beta_0 < 0$. More importantly, it can be considered as a specific expression for the modification of the original Friedmann equation by the higher order parameter $\beta_0$, which will be used in subsequent studies.  When $\beta_0=0$, Eq.~(\ref{eq16}) returns to the case of original standard cosmological model, that is $Z_{\beta_0}  = 1$. Furthermore, for the convenience of the ensuing discussion, we rewrite Eq.~(\ref{eq10}) as follows
\begin{align}
\label{eq16+}
H_{\beta_0}=H_0 + \delta H,
\end{align}
where $\delta H = \left( {{Z_{{\beta _0}}} - 1} \right){H_0}$.

Next, we explore the impact of GUP on BBN. According to the standard BBN model, neutron $n$ and
protons $p$ began to form only few thousandths of a second after the Big Bang, when the temperature dropped low enough to reach the freeze-out temperature. The abundances of the initial very light atomic nuclei were defined from the first hundredth of a second to a few minutes. As is well known, shortly after the Big Bang, our universe was mainly composed of hydrogen and helium, as well as small amounts of lithium and beryllium. In particular, the formation of the primordial ${^4} He$ occurs at the temperature $T$ around $100\text{MeV}$, when the energy and number density were dominated by relativistic leptons (electrons $e^-$, positrons $e^+$, neutrinos $\nu$) and photons $\gamma$ (neutron $n$ and protons $p$).  These particles are in thermal equilibrium due to the fast rate of their collisions, which makes ${T_\nu } = {T_e} = {T_\gamma } = T$~\cite{Bernstein:1988ad}.

On the other hand, the scattering of neutrons $n$ and protons $p$ is maintained in equilibrium and does not contribute to the total energy density. In thermal equilibrium, weak reaction processes allow for the inter-conversion between neutrons $n$ and protons $p$~\cite{Suh:1999va}:
\begin{subequations}
\label{eq17}
\begin{align}
\textbf{I}:  {\nu } + n \leftrightarrow p + {e^ - },
\\
\textbf{II}:  {e^ + } + n \leftrightarrow p + {\nu },
\\
\textbf{III}:  n \leftrightarrow p + {e^ - } + {\nu }.
\end{align}
\end{subequations}
The Eq.~(\ref{eq17}) implies that the neutron abundance in the expanding universe can be estimated by calculating  the conversion rate ${\lambda _{pn}}\left(T \right)$ of protons into neutrons and its inverse ${\lambda _{np}}\left( T \right)$.  In turn, at sufficiently high temperatures, the total weak interaction rate can be expressed in terms of ${\lambda _{pn}}\left( T \right)$ and ${\lambda _{np}}\left( T \right)$, i.e.
\begin{align}
\label{eq18}
{\rm{\Lambda }}\left( T \right) = {\lambda _{np}}\left( T \right) + {\lambda _{pn}}\left( T \right),
\end{align}
where ${\lambda _{np}}\left( T \right) = {\lambda _\text{I}}\left( T \right) + {\lambda _\text{II}}\left( T \right) + {\lambda _\text{III}}\left( T \right)$ is represented as the sum of the rates associated with each of the three processes in Eq.~(\ref{eq17}), respectively. Moreover, the quantities ${\lambda _{pn}}\left(T \right)$  and ${\lambda _{np}}\left( T \right)$ are satisfying the relationship ${\lambda _{np}}\left( T \right) = \exp \left( { - {\cal Q}/T} \right){\lambda _{pn}}\left( T \right)$ where ${\cal Q} = {m_n} - {m_p}$ characterizes the difference between neutron and proton masses.

In Ref.~\cite{Bernstein:1988ad}, it is pointed out that during the freeze-out period, the temperature $T$ is lower compared to the characteristic energy $E$ that contributes to the rates for the decays Eq.~(\ref{eq17}). Therefore, one can estimate the lepton phase-space density functions by the ``classical-Boltzmann weights, rather than the Fermi-Dirac distribution. Furthermore, by assuming that the mass of electrons relative to the energy of electrons and neutrinos can be negligible, the total sum of the weak interaction can be repressed as \cite{Bernstein:1988ad,Ghoshal:2021ief}
\begin{align}
\label{eq19}
{{\Lambda }}\left( T \right) & = 4A{{T}^{3}}(4!{{T}^{2}}+2\times 3! {\cal Q} T+2!{{\cal Q}^{2}})
 \nonumber \\
& \simeq q{T^5} + {\cal O}\left( {\frac{{\cal Q}}{T}} \right),
\end{align}
where $A = 1.02 \times {10^{ - 11}}{\text{Ge}}{{\text{V}}^{ - 4}}$ and $q = 9.6 \times {10^{ - 10}}{\text{Ge}}{{\text{V}}^{ - 4}}$. The primordial  mass fraction of $^4 He$ can be estimated as \cite{Ghoshal:2021ief,Kolb:1981hk}
\begin{align}
\label{eq20}
{Y_p} = \gamma \frac{{2x\left( {{t_f}} \right)}}{{1 + x\left( {{t_f}} \right)}},
\end{align}
where $\gamma  = \exp \left[ {{{ - \left( {{t_n} - {t_f}} \right)} \mathord{\left/ {\vphantom {{ - \left( {{t_n} - {t_f}} \right)} \tau }} \right.
\kern-\nulldelimiterspace} \tau }} \right] \simeq 1$, where $t_f$ and $t_n$ are the the freeze-out of the weak interactions and the time of the freeze-out of the nucleosynthesis, respectively. The neutron mean lifetime $\tau$ is taken as $877$s.
The function $\gamma \left( {{t_f}} \right)$ is defined as the fraction of neutrons that decay into protons during the interval $t \in \left[ {{t_f},{t_n}} \right]$, and $x\left( {{t_f}} \right) = \exp \left[ {{{ - \cal{Q}} \mathord{\left/
 {\vphantom {{ - \cal{Q}} {T\left( {{t_f}} \right)}}} \right. \kern-\nulldelimiterspace} {T\left( {{t_f}} \right)}}} \right]$ is the neutron-to-proton equilibrium ratio. The deviations in the fractional mass $Y_p$ resulting from the variation of the freeze-out temperature $T_f$ can be expressed as follows \cite{Ghoshal:2021ief}
\begin{align}
\label{eq21}
\delta {Y_p} = {Y_p}\left[ {\left( {1 - \frac{{{Y_p}}}{{2\gamma }}} \right)\log \left( {\frac{{2\gamma }}{{{Y_p}}} - 1} \right) - \frac{{2{t_f}}}{\tau }} \right]\frac{{\delta {T_f}}}{{{T_f}}}.
\end{align}
Notably, $T_n$ is fixed by the deuterium binding energy, so that one can set $\delta {Y_n}\left( {{t_n}} \right)=0$~\cite{Capozziello:2017bxm,Lambiase:2012fv}. In Ref.~\cite{Fields:2011zzb}, the authors utilized both visible and infrared $^4 He$ emission lines in  extragalactic HII regions to determine the mass fraction of $^4 He$, which shows the fractional mass as follows
\begin{align}
\label{eq22}
{Y_p} = 0.2449, \quad  \left| {\delta {Y_p}} \right| \lesssim {10^{ - 4}}.
\end{align}
By substituting these astronomical observations into Eq.~(\ref{eq21}), then setting ${t_f} \simeq 1 \text{s}$ and ${t_n} \simeq 20 \text{s}$, one has
\begin{align}
\label{eq23}
\left| {\frac{{\delta {T_f}}}{{{T_f}}}} \right| \lesssim {4.7\times 10^{ - 4}},
\end{align}

According to Refs.~\cite{Luciano:2021vkl,Ghoshal:2021ief,Capozziello:2017bxm}, one has
the relation $\Lambda = H$, allows to derive the freeze-out temperature $T = {T_f}\left( {1 + {{\delta {T_f}} \mathord{\left/  {\vphantom {{\delta {T_f}} {{T_f}}}} \right.  \kern-\nulldelimiterspace} {{T_f}}}} \right)$, with ${T_{f}}\sim 0.6 \text{MeV}$. Then, according to Eq.~(\ref{eq16+}) and Eq.~(\ref{eq19}), it is found that $\delta \Lambda  = 5q{T_f}^4\delta T =\delta H$ \cite{Capozziello:2017bxm}, which giving
\begin{align}
\label{eq21+}
\frac{{\delta T}}{{{T_f}}} = \left( {{Z_{{\beta _0}}} - 1} \right)\frac{{{H_0}}}{{5qT_f^5}}
\end{align}
By substituting Eq.~(\ref{eq15}) and Eq.~(\ref{eq16}) into Eq.~(\ref{eq21+}), one has
\begin{subequations}
\label{22}
\begin{align}
{\left. {\frac{{\delta {T_f}}}{{{T_f}}}} \right|_{{\beta _0} > 0}} =  \frac{{8g{G^2}\sqrt {{g_*}G} {\pi ^{9/2}}{\beta _0}{T_f}}}{{675\sqrt 5 q}},
\\
{\left. {\frac{{\delta {T_f}}}{{{T_f}}}} \right|_{{\beta _0} < 0}}= - \frac{{8g{G^2}\sqrt {{g_*}G} {\pi ^{9/2}}{\beta _0}{T_f}}}{{675\sqrt 5 q}}.
\end{align}
\end{subequations}
Combining  Eq.~(\ref{eq23}), the bounds of the GUP parameter can be obtained. By a simple calculation, the result is given by
\begin{align}
\label{}
- 2.5 \times {10^{84}} \lesssim {\beta _0}\lesssim 2.5 \times {10^{84}}
\end{align}
Now, based on the variation of the freeze-out temperature, the bounds of parameter $\beta_0$ are obtained. Compared to previous work, our results dose not only give the upper bound of $\beta_0$, but also the lower bound of $\beta_0$, which suggests that the higher-order GUP model can show more physical information than the KMM model . More importantly, in contrast to Ref.~\cite{Luciano:2021vkl}, our GUP model has stronger constraints on the parameter $\beta_0$. In a word, our results also show that GUP models with both positive and negative parameters can have an impact on BBN cosmology.

\section{Primordial light element abundances($^4 He, D, ^7 Li$) constrain GUP parameter $\beta_0$}
\label{sec4}
In this section, we will constrain the bounds of GUP parameter $\beta_0$ by requiring consistency between GUP cosmology predictions and observational data of the primordial light element abundances, such as Helium isotope $^4 He$, Deuterium $D$, and Lithium
isotope $^7 Li$~\cite{Luciano:2021vkl,Ghoshal:2021ief}. In order to achieve this goal, one needs to study the deviation of primordial light element abundance from standard cosmology caused by GUP correction. This research scenario requires us to replace the standard $Z$-factor of the primordial abundance with a $Z$-function related to $\beta_0$, which is related to the effective number of neutrino species~\cite{Boran:2013pux}. In the classical theory of cosmology, the original $Z$-factor equals 1.

The deviation of $Z$ from unit may be caused by two situations: one is the modification of gravity, another is  the presence of additional light particles (such as neutrinos), in which case one has
\begin{align}
\label{eq26}
{Z_\nu } = {\left[ {1 + \frac{7}{{43}}\left( {{N_\nu } - 3} \right)} \right]^{\frac{1}{2}}},
\end{align}
where ${N_\nu }$ is the number of neutrino generations. Considering that our goal is the effect of GUP on BBN, we further assume $N = 3$ in order to exclude the possibility that the $Z$ deviation is caused by the degrees of freedom of the additional particles.

\subsection{$^4 He$ abundance}
Follow the viewpoint in Refs.~\cite{Ghoshal:2021ief,Boran:2013pux}, the Helium isotope $^4 He$ can be produced by the following nuclear reactions. First, deuterium $D$ is produced from a neutron $n$ and a proton $p$, and then the $D$ is converted into $^3 {He}$ and tritium $^3{H}$. The relevant reactions are
\begin{subequations}
\begin{align}
\label{eq27}
n + p & \to D + \gamma,
\\
D + D & \to {^3 He} + {{n}},
\\
D + D  & \to {^3 H} + p,
\end{align}
\end{subequations}
The final step leads to the production of $^4 He$ due to the following processes
\begin{subequations}
\begin{align}
\label{eq28}
D + {^3 H} \to {^4 He} + {{n}},
\\
D + {^3 He} \to {^4 He} + p.
\end{align}
\end{subequations}
Based on the results in Refs.~\cite{Kneller:2004jz,Steigman:2007xt}, the numerical best fit constrains the primordial $^4 He$ abundance to be
\begin{align}
\label{eq29}
{Y_p} = 0.2485 + 0.0016\left[ {\left( {{\eta _{10}} - 6} \right) + 100\left( {Z - 1} \right)} \right].
\end{align}
%In the above equation, the ``$+$" represents the case where $\beta_0>0$, while the ``$-$" denotes the case where $\beta_0<0$.
In this work, we have $Z=Z_{\beta_0}$ given by Eq.~(\ref{eq16}). The baryon density parameter is usually defined as~\cite{Steigman:2012ve,Simha:2008zj,Schoneberg:2024ifp}
\begin{align}
\label{eq30}
{\eta _{10}} \equiv {10^{10}}{\eta _B} \equiv {10^{10}}{{{n_B}} \mathord{\left/ {\vphantom {{{n_B}} {{n_\gamma }}}} \right.
 \kern-\nulldelimiterspace} {{n_\gamma }}},
\end{align}
where ${{{n_B}} \mathord{\left/ {\vphantom {{{n_B}} {{n_\gamma }}}} \right.
 \kern-\nulldelimiterspace} {{n_\gamma }}}$ represents the ratio of baryons to photons~\cite{WMAP:2010qai}. For $Z=1$, the BBN model predicts the primordial $^4 He$ abundance ${Y_p}{|_{\text{GR}}} = 0.2485 $. However, based on observational data~\cite{Ghoshal:2021ief,Fields:2019pfx,Bhattacharjee:2020uhs} and settings ${\eta _{10}} \simeq 6$, the $^4 He$ abundance is $0.2449$. In order to maintain consistency between the observed data and the numerical fit~(\ref{eq29}), one has
\begin{align}
\label{eq31}
0.2449   = 0.2485 + 0.016\left[ {100\left( {Z - 1} \right)} \right].
\end{align}
Therefore, the value of $Z$ is given by
\begin{align}
\label{eq32}
Z = 1.0475 .
\end{align}
Now, replacing the $Z$-factor with a $Z$-function(\ref{eq16}) , the deviation of the $Z$ function from the standard can be described as
\begin{align}
\label{eq32.1}
\delta Z & \equiv Z - 1
\nonumber \\
& =  \left( {1 \pm \frac{4}{{45}}g{G^2}{\pi ^3}{T^4}{\beta _0}} \right) - 1 \lesssim 0.0475.
\end{align}
In the above equation, the positive sign represents $\beta_0 >0$, while negative sign means $\beta_0 <0 $. In order to explore the constraints of deviation from standard cosmology on GUP parameter, we use the expression of $Z_{\beta_0}$ and set the temperature $T \simeq 10 \text{MeV}$ to obtain the bound of GUP parameter $\beta_0$. For $\beta_0>0$, the bound of GUP parameter is
\begin{align}
\label{eq33}
\beta_0  \lesssim 1.72 \times {10^{81}}.
\end{align}
With the same method, one can obtain the results for the case $\beta_0<0$ as follow
\begin{align}
\label{eq34}
\beta_0  \gtrsim -1.72 \times {10^{81}}.
\end{align}
It can be seen that our results give both upper and lower bounds on $\beta_0$, which does not appear in previous works~\cite{Das:2021nbq,Luciano:2021vkl,Kouwn:2018rmp,Giardino:2020myz}.

\subsection{$D$ abundance}
According to the analysis of the primordial $^4 He$ abundance, the process of $D$ generation is represented as $n + p \to D + \gamma $. Determining the  primordial abundance of $D$ through numerical fitting~\cite{Steigman:2012ve}, one has
\begin{align}
\label{eq35}
{y_{{D_p}}} = 2.6{\left[ {\frac{6}{{{\eta _{10}} - 6\left( {Z - 1} \right)}}} \right]^2}.
\end{align}
As previously stated, the values $Z = 1$ and ${\eta _{10}} = 6$ yields the standard result in GR $\left( {{y_{{D_p}}}} \right){|_{\text{GR}}} = 2.6$. According to the observational data~\cite{Fields:2019pfx,Ghoshal:2021ief,Bhattacharjee:2020uhs}, the $D$ abundance constraint is ${y_{{D_p}}} = 2.55$, combined with Eq.~(\ref{eq35}), there are
\begin{align}
\label{eq36}
Z = 1.062.
\end{align}
It is found that the bounds of the  primordial $D$ abundance on the $Z$-function partially overlaps with $^4 He$. Limiting the range of the $Z$-function to $\delta Z \equiv Z-1 \lesssim 0.062$ with $Z=Z_{\beta_0}$, one has
\begin{align}
\label{eq36+}
\delta Z = \left(1 \pm \frac{4}{{45}}g{G^2}{\pi ^3}{T^4}{\beta _0} \right) - 1 \lesssim 0.062.
\end{align}
According to the above equation and setting temperature $T \simeq 10\rm{MeV}$, the bounds of GUP parameter $\beta_0$ can be  constrained, and the results are as follows
\begin{align}
\label{eq37}
-2.24\times {10^{81}} \lesssim \beta_0  \lesssim 2.24 \times {10^{81}}.
\end{align}
Obviously, the above equation now gives not only the upper but also the lower bound of the GUP parameter.

\subsection{$^7 Li$ abundance}
As is well known, the baryon density parameter ${\eta _{10}} = 6$ fully conforms to the abundance of $^4 He$, $D$, and other light elements, but is inconsistent with the observed results of $^7 Li$. In fact, according to standard cosmological theory~\cite{Fields:2011zzb,Boran:2013pux,Luciano:2022knb}, the ratio of $^7 Li$ abundance range from $\text{2.4}$ to $\text{4.3}$. However, currently neither the standard BBN model nor any modified cosmological model can fit such low abundance ratios, a challenge known as the $Li$ problem. In this work, in order to address the issue of  primordial $Li$ abundance, the GUP parameters are constrained by deviating from the standard cosmological $Z$-function as with the previous method. The optimal expression for the primordial $Li$ abundance through numerical fitting is given by is~\cite{Steigman:2012ve}
\begin{align}
\label{eq39}
{y_{Li}} = 4.82{\left( {\frac{{{\eta _{10}} - 3\left( {Z - 1} \right)}}{6}} \right)^2},
\end{align}

Here, based on the findings in Refs.~\cite{Ghoshal:2021ief,Fields:2019pfx,Steigman:2012ve}, the observational constraint on Lithium abundance is ${y_{Li}} = 1.6 $. Moreover, by setting ${\eta _{10}} = 6$, one can constrain deviations of $Z$ from unity by demanding consistency between the numerical best fit expression. This leads to the constraint on $Z = 1.960025 $, which is obviously different from $^4 He$ and $D$. Using $Z=Z_{\beta_0}$, the range of the $Z$-function can be restricted to  $\delta Z \equiv Z-1 \lesssim 0.960025$. which leads to
\begin{align}
\label{eq39+}
\delta Z = \left( 1 \pm \frac{4}{{45}}g{G^2}{\pi ^3}{T^4}{\beta _0} \right) - 1 \lesssim 0.960025.
\end{align}
By setting the temperature $T \simeq 10 \text{MeV}$, the upper bound of GUP parameter is given by
\begin{align}
\label{eq40}
\beta_0 \lesssim 3.48 \times {10^{82}},
\end{align}
which is for the parameters $\beta_0>0$. When the parameter $\beta_0< 0$, similarly, it can be obtained that
\begin{align}
\label{eq41}
\beta_0  \gtrsim  - 3.48 \times {10^{82}}.
\end{align}
One can see that the parameter range of $^7 Li$ abundance partially overlaps with the parameter bounds of $^4 He$, $D$, and the bounds are greater than the other two light elements. In addition, we also investigated the negative parameter GUP model and obtained the bounds when the parameter takes a negative sign. Therefore, we speculate that selecting appropriate parameters may be one of the methods to solve the  $Li$ problem. This will be an interesting study that requires further attention and will be conducted elsewhere.

\section{Conclusion}
\label{sec5}
In this paper, we investigate the primordial BBN of the universe and the related abundances of primordial light elements in the framework of a new higher-order GUP. To begin with, we corrected the Bekenstein-Hawking entropy using the new higher-order GUP model, then combined the corrected entropy with the first law of thermodynamics to derive the modified Friedmann equations, as shown in Eq.~(\ref{eq9}). Subsequently, these modified Friedmann equations were used to analyze the BBN. In addition, incorporating our study with astronomical observations, we constrained the range of the GUP parameters. The results show the bounds on GUP parameter $\beta_0$ are $- {{10}^{84}}$ to ${{10}^{84}}$. Consistency with observational data of primordial abundances of $^4 He, D, ^7 Li$, we obtain the various
constraints on the GUP parameter $\beta_0$. Comparison with previous works, it is found that the upper bounds of the GUP parameters in this paper are similar to those obtained in previous work using cosmological analysis~\cite{Kouwn:2018rmp,Giardino:2020myz,Gine:2020izd}, while lower than the weaker quantum gravity experiments~\cite{Ali:2011fa,Ghosh:2013qra,Bushev:2019zvw}. However, it is important to note that our results also give a lower bounds on the GUP parameter, which could not be given by previous works.  This indicates that the GUP parameter bounds are not limited to the positive domain but also extend to the negative domain. Therefore, the GUP model with negative parameters can influence the classical system of physics.

Finally, it is interesting to discuss the impact of extended uncertainty principle (EUP) on cosmology. The EUP is another correction of Heisenberg's uncertainty principle considers a large-scale correction, known as the extended uncertainty principle (EUP), which takes into account a fundamental length scale $L_*$~\cite{Bambi:2007ty,Kempf:1994su,Mureika:2018gxl}. The EUP correction has only recently received attention since it was previously believed that there is no need for large-scale corrections to gravitational physics. However, the current situation has changed,  EUP provides a way to introduce quantum effects at macroscopic distances, which can be applied to physical systems at large distance scales and in non-extreme cases and obtained the lower bounds for the EUP fundamental length scale $L_*$~\cite{Okcu:2022sio}. In recent years, EUP has been used to modify the FRW cosmology and (anti) de Sitter space thermodynamics~\cite{Moradpour:2019yiq,Chung:2019iwp,Dabrowski:2019wjk}. As is known to all, the formation of primordial ${^4}He$ occurs at temperatures around  ${\rm{T}} \simeq 100{\rm{MeV}}$, which corresponds to approximately $10$ seconds after the Big Bang. At this point, the universe can be considered in an non-extreme case. Considering that EUP is a large-scale correction to gravitational physics, we argue that the EUP model may be more suitable for investigating primordial BBN. As mentioned above, a further direction to explore is the study of effects of EUP modles on BBN cosmological model, which provides a deep insight into the understanding QG and  large length scale corrections to the dynamics of the FRW universe.

\bibliography{example}

%Merlin.mbs v4.21 2009-07-09.
\begin{thebibliography}{10}%
\makeatletter
\providecommand \@ifxundefined [1]{%
 \ifx #1\undefined \expandafter \@firstoftwo
 \else \expandafter \@secondoftwo
\fi
}%
\providecommand \@ifnum [1]{%
 \ifnum #1\expandafter \@firstoftwo
 \else \expandafter \@secondoftwo
\fi
}%
\providecommand \enquote [1]{``#1''}%
\providecommand \bibnamefont  [1]{#1}%
\providecommand \bibfnamefont [1]{#1}%
\providecommand \citenamefont [1]{#1}%
\providecommand\href[0]{\@sanitize\@href}%
\providecommand\@href[1]{\endgroup\@@startlink{#1}\endgroup\@@href}%
\providecommand\@@href[1]{#1\@@endlink}%
\providecommand \@sanitize [0]{\begingroup\catcode`\&12\catcode`\#12\relax}%
\@ifxundefined \pdfoutput {\@firstoftwo}{%
 \@ifnum{\z@=\pdfoutput}{\@firstoftwo}{\@secondoftwo}%
}{%
 \providecommand\@@startlink[1]{\leavevmode\special{html:<a href="#1">}}%
 \providecommand\@@endlink[0]{\special{html:</a>}}%
}{%
 \providecommand\@@startlink[1]{%
  \leavevmode
  \pdfstartlink
   attr{/Border[0 0 1 ]/H/I/C[0 1 1]}%
   user{/Subtype/Link/A<</Type/Action/S/URI/URI(#1)>>}%
  \relax
 }%
 \providecommand\@@endlink[0]{\pdfendlink}%
}%
\providecommand \url  [0]{\begingroup\@sanitize \@url }%
\providecommand \@url [1]{\endgroup\@href {#1}{\urlprefix}}%
\providecommand \urlprefix [0]{URL }%
\providecommand \Eprint[0]{\href }%
\@ifxundefined \urlstyle {%
  \providecommand \doi [1]{doi:\discretionary{}{}{}#1}%
}{%
  \providecommand \doi [0]{doi:\discretionary{}{}{}\begingroup
  \urlstyle{rm}\Url }%
}%
\providecommand \doibase [0]{http://dx.doi.org/}%
\providecommand \Doi[1]{\href{\doibase#1}}%
\providecommand \bibAnnote [3]{%
  \BibitemShut{#1}%
  \begin{quotation}\noindent
    \textsc{Key:}\ #2\\\textsc{Annotation:}\ #3%
  \end{quotation}%
}%
\providecommand \bibAnnoteFile [2]{%
  \IfFileExists{#2}{\bibAnnote {#1} {#2} {\input{#2}}}{}%
}%
\providecommand \typeout [0]{\immediate \write \m@ne }%
\providecommand \selectlanguage [0]{\@gobble}%
\providecommand \bibinfo [0]{\@secondoftwo}%
\providecommand \bibfield [0]{\@secondoftwo}%
\providecommand \translation [1]{[#1]}%
\providecommand \BibitemOpen[0]{}%
\providecommand \bibitemStop [0]{}%
\providecommand \bibitemNoStop [0]{.\EOS\space}%
\providecommand \EOS [0]{\spacefactor3000\relax}%
\providecommand \BibitemShut [1]{\csname bibitem#1\endcsname}%
%</preamble>
\bibitem{Zeilinger}%
  \BibitemOpen
  \bibfield{author}{%
  \bibinfo {author} {\bibfnamefont{A.}~\bibnamefont{Zeilinger}},\ }%
  \bibfield{journal}{%
  \Doi{10.1103/RevModPhys.71.S288}{\bibinfo {journal} {Rev. Mod. Phys.}}\ }%
  \textbf{\bibinfo {volume} {71}},\ \bibinfo {pages} {S288} (\bibinfo {year}
  {1999})%
  \bibAnnoteFile{NoStop}{Zeilinger}%
\bibitem{Will:2014kxa}%
  \BibitemOpen
  \bibfield{author}{%
  \bibinfo {author} {\bibfnamefont{C.~M.}\ \bibnamefont{Will}},\ }%
  \bibfield{journal}{%
  \Doi{10.12942/lrr-2014-4}{\bibinfo {journal} {Living Rev. Rel.}}\ }%
  \textbf{\bibinfo {volume} {17}},\ \bibinfo {pages} {4} (\bibinfo {year}
  {2014}),\ \Eprint{http://arxiv.org/abs/1403.7377}{arXiv:1403.7377 [gr-qc]}%
  \bibAnnoteFile{NoStop}{Will:2014kxa}%
\bibitem{Scardigli:1999jh}%
  \BibitemOpen
  \bibfield{author}{%
  \bibinfo {author} {\bibfnamefont{F.}~\bibnamefont{Scardigli}},\ }%
  \bibfield{journal}{%
  \Doi{10.1016/S0370-2693(99)00167-7}{\bibinfo {journal} {Phys. Lett. B}}\ }%
  \textbf{\bibinfo {volume} {452}},\ \bibinfo {pages} {39} (\bibinfo {year}
  {1999}),\ \Eprint{http://arxiv.org/abs/hep-th/9904025}{arXiv:hep-th/9904025}%
  \bibAnnoteFile{NoStop}{Scardigli:1999jh}%
\bibitem{Maggiore:1993kv}%
  \BibitemOpen
  \bibfield{author}{%
  \bibinfo {author} {\bibfnamefont{M.}~\bibnamefont{Maggiore}},\ }%
  \bibfield{journal}{%
  \Doi{10.1016/0370-2693(93)90785-G}{\bibinfo {journal} {Phys. Lett. B}}\ }%
  \textbf{\bibinfo {volume} {319}},\ \bibinfo {pages} {83} (\bibinfo {year}
  {1993}),\ \Eprint{http://arxiv.org/abs/hep-th/9309034}{arXiv:hep-th/9309034}%
  \bibAnnoteFile{NoStop}{Maggiore:1993kv}%
\bibitem{Amelino-Camelia:2000stu}%
  \BibitemOpen
  \bibfield{author}{%
  \bibinfo {author} {\bibfnamefont{G.}~\bibnamefont{Amelino-Camelia}},\ }%
  \bibfield{journal}{%
  \Doi{10.1142/S0218271802001330}{\bibinfo {journal} {Int. J. Mod. Phys. D}}\
  }%
  \textbf{\bibinfo {volume} {11}},\ \bibinfo {pages} {35} (\bibinfo {year}
  {2002}),\ \Eprint{http://arxiv.org/abs/gr-qc/0012051}{arXiv:gr-qc/0012051}%
  \bibAnnoteFile{NoStop}{Amelino-Camelia:2000stu}%
\bibitem{Garay:1994en}%
  \BibitemOpen
  \bibfield{author}{%
  \bibinfo {author} {\bibfnamefont{L.~J.}\ \bibnamefont{Garay}},\ }%
  \bibfield{journal}{%
  \Doi{10.1142/S0217751X95000085}{\bibinfo {journal} {Int. J. Mod. Phys. A}}\
  }%
  \textbf{\bibinfo {volume} {10}},\ \bibinfo {pages} {145} (\bibinfo {year}
  {1995}),\ \Eprint{http://arxiv.org/abs/gr-qc/9403008}{arXiv:gr-qc/9403008}%
  \bibAnnoteFile{NoStop}{Garay:1994en}%
\bibitem{Medved:2004yu}%
  \BibitemOpen
  \bibfield{author}{%
  \bibinfo {author} {\bibfnamefont{A.~J.~M.}\ \bibnamefont{Medved}}\ and\
  \bibinfo {author} {\bibfnamefont{E.~C.}\ \bibnamefont{Vagenas}},\ }%
  \bibfield{journal}{%
  \Doi{10.1103/PhysRevD.70.124021}{\bibinfo {journal} {Phys. Rev. D}}\ }%
  \textbf{\bibinfo {volume} {70}},\ \bibinfo {pages} {124021} (\bibinfo {year}
  {2004}),\ \Eprint{http://arxiv.org/abs/hep-th/0411022}{arXiv:hep-th/0411022}%
  \bibAnnoteFile{NoStop}{Medved:2004yu}%
\bibitem{Amelino-Camelia:2004uiy}%
  \BibitemOpen
  \bibfield{author}{%
  \bibinfo {author} {\bibfnamefont{G.}~\bibnamefont{Amelino-Camelia}}, \bibinfo
  {author} {\bibfnamefont{M.}~\bibnamefont{Arzano}},\ and\ \bibinfo {author}
  {\bibfnamefont{A.}~\bibnamefont{Procaccini}},\ }%
  \bibfield{journal}{%
  \Doi{10.1103/PhysRevD.70.107501}{\bibinfo {journal} {Phys. Rev. D}}\ }%
  \textbf{\bibinfo {volume} {70}},\ \bibinfo {pages} {107501} (\bibinfo {year}
  {2004}),\ \Eprint{http://arxiv.org/abs/gr-qc/0405084}{arXiv:gr-qc/0405084}%
  \bibAnnoteFile{NoStop}{Amelino-Camelia:2004uiy}%
\bibitem{Khodadi:2017eim}%
  \BibitemOpen
  \bibfield{author}{%
  \bibinfo {author} {\bibfnamefont{M.}~\bibnamefont{Khodadi}}, \bibinfo
  {author} {\bibfnamefont{K.}~\bibnamefont{Nozari}},\ and\ \bibinfo {author}
  {\bibfnamefont{A.}~\bibnamefont{Hajizadeh}},\ }%
  \bibfield{journal}{%
  \Doi{10.1016/j.physletb.2017.05.016}{\bibinfo {journal} {Phys. Lett. B}}\ }%
  \textbf{\bibinfo {volume} {770}},\ \bibinfo {pages} {556} (\bibinfo {year}
  {2017}),\ \Eprint{http://arxiv.org/abs/1702.06357}{arXiv:1702.06357 [gr-qc]}%
  \bibAnnoteFile{NoStop}{Khodadi:2017eim}%
\bibitem{Chemissany:2011nq}%
  \BibitemOpen
  \bibfield{author}{%
  \bibinfo {author} {\bibfnamefont{W.}~\bibnamefont{Chemissany}}, \bibinfo
  {author} {\bibfnamefont{S.}~\bibnamefont{Das}}, \bibinfo {author}
  {\bibfnamefont{A.~F.}\ \bibnamefont{Ali}},\ and\ \bibinfo {author}
  {\bibfnamefont{E.~C.}\ \bibnamefont{Vagenas}},\ }%
  \bibfield{journal}{%
  \Doi{10.1088/1475-7516/2011/12/017}{\bibinfo {journal} {JCAP}}\ }%
  \textbf{\bibinfo {volume} {12}},\ \bibinfo {pages} {017} (\bibinfo {year}
  {2011}),\ \Eprint{http://arxiv.org/abs/1111.7288}{arXiv:1111.7288}%
  \bibAnnoteFile{NoStop}{Chemissany:2011nq}%
\bibitem{Chen:2014jwq}%
  \BibitemOpen
  \bibfield{author}{%
  \bibinfo {author} {\bibfnamefont{P.}~\bibnamefont{Chen}}, \bibinfo {author}
  {\bibfnamefont{Y.~C.}\ \bibnamefont{Ong}},\ and\ \bibinfo {author}
  {\bibfnamefont{D.-h.}\ \bibnamefont{Yeom}},\ }%
  \bibfield{journal}{%
  \Doi{10.1016/j.physrep.2015.10.007}{\bibinfo {journal} {Phys. Rept.}}\ }%
  \textbf{\bibinfo {volume} {603}},\ \bibinfo {pages} {1} (\bibinfo {year}
  {2015}),\ \Eprint{http://arxiv.org/abs/1412.8366}{arXiv:1412.8366}%
  \bibAnnoteFile{NoStop}{Chen:2014jwq}%
\bibitem{Moradpour:2019wpj}%
  \BibitemOpen
  \bibfield{author}{%
  \bibinfo {author} {\bibfnamefont{H.}~\bibnamefont{Moradpour}}, \bibinfo
  {author} {\bibfnamefont{A.~H.}\ \bibnamefont{Ziaie}}, \bibinfo {author}
  {\bibfnamefont{S.}~\bibnamefont{Ghaffari}},\ and\ \bibinfo {author}
  {\bibfnamefont{F.}~\bibnamefont{Feleppa}},\ }%
  \bibfield{journal}{%
  \Doi{10.1093/mnrasl/slz098}{\bibinfo {journal} {Mon. Not. Roy. Astron.
  Soc.}}\ }%
  \textbf{\bibinfo {volume} {488}},\ \bibinfo {pages} {L69} (\bibinfo {year}
  {2019}),\ \Eprint{http://arxiv.org/abs/1907.12940}{arXiv:1907.12940}%
  \bibAnnoteFile{NoStop}{Moradpour:2019wpj}%
\bibitem{Feng:2015jlj}%
  \BibitemOpen
  \bibfield{author}{%
  \bibinfo {author} {\bibfnamefont{Z.~W.}\ \bibnamefont{Feng}}, \bibinfo
  {author} {\bibfnamefont{H.~L.}\ \bibnamefont{Li}}, \bibinfo {author}
  {\bibfnamefont{X.~T.}\ \bibnamefont{Zu}},\ and\ \bibinfo {author}
  {\bibfnamefont{S.~Z.}\ \bibnamefont{Yang}},\ }%
  \bibfield{journal}{%
  \Doi{10.1140/epjc/s10052-016-4057-1}{\bibinfo {journal} {Eur. Phys. J. C}}\
  }%
  \textbf{\bibinfo {volume} {76}},\ \bibinfo {pages} {212} (\bibinfo {year}
  {2016}),\ \Eprint{http://arxiv.org/abs/1604.04702}{arXiv:1604.04702}%
  \bibAnnoteFile{NoStop}{Feng:2015jlj}%
\bibitem{Casadio:2020rsj}%
  \BibitemOpen
  \bibfield{author}{%
  \bibinfo {author} {\bibfnamefont{R.}~\bibnamefont{Casadio}}\ and\ \bibinfo
  {author} {\bibfnamefont{F.}~\bibnamefont{Scardigli}},\ }%
  \bibfield{journal}{%
  \Doi{10.1016/j.physletb.2020.135558}{\bibinfo {journal} {Phys. Lett. B}}\ }%
  \textbf{\bibinfo {volume} {807}},\ \bibinfo {pages} {135558} (\bibinfo {year}
  {2020}),\ \Eprint{http://arxiv.org/abs/2004.04076}{arXiv:2004.04076}%
  \bibAnnoteFile{NoStop}{Casadio:2020rsj}%
\bibitem{Iorio:2019wtn}%
  \BibitemOpen
  \bibfield{author}{%
  \bibinfo {author} {\bibfnamefont{A.}~\bibnamefont{Iorio}}, \bibinfo {author}
  {\bibfnamefont{G.}~\bibnamefont{Lambiase}}, \bibinfo {author}
  {\bibfnamefont{P.}~\bibnamefont{Pais}},\ and\ \bibinfo {author}
  {\bibfnamefont{F.}~\bibnamefont{Scardigli}},\ }%
  \bibfield{journal}{%
  \Doi{10.1103/PhysRevD.101.105002}{\bibinfo {journal} {Phys. Rev. D}}\ }%
  \textbf{\bibinfo {volume} {101}},\ \bibinfo {pages} {105002} (\bibinfo {year}
  {2020}),\ \Eprint{http://arxiv.org/abs/1910.09019}{arXiv:1910.09019}%
  \bibAnnoteFile{NoStop}{Iorio:2019wtn}%
\bibitem{Park:2020zom}%
  \BibitemOpen
  \bibfield{author}{%
  \bibinfo {author} {\bibfnamefont{D.}~\bibnamefont{Park}}\ and\ \bibinfo
  {author} {\bibfnamefont{E.}~\bibnamefont{Jung}},\ }%
  \bibfield{journal}{%
  \Doi{10.1103/PhysRevD.101.066007}{\bibinfo {journal} {Phys. Rev. D}}\ }%
  \textbf{\bibinfo {volume} {101}},\ \bibinfo {pages} {066007} (\bibinfo {year}
  {2020}),\ \Eprint{http://arxiv.org/abs/2001.02850}{arXiv:2001.02850}%
  \bibAnnoteFile{NoStop}{Park:2020zom}%
\bibitem{Battista:2024gud}%
  \BibitemOpen
  \bibfield{author}{%
  \bibinfo {author} {\bibfnamefont{E.}~\bibnamefont{Battista}}, \bibinfo
  {author} {\bibfnamefont{S.}~\bibnamefont{Capozziello}},\ and\ \bibinfo
  {author} {\bibfnamefont{A.}~\bibnamefont{Errehymy}}\ }%
  \Eprint{http://arxiv.org/abs/2409.09750}{arXiv:2409.09750}%
  \bibAnnoteFile{NoStop}{Battista:2024gud}%
\bibitem{Xiang:2009yq}%
  \BibitemOpen
  \bibfield{author}{%
  \bibinfo {author} {\bibfnamefont{L.}~\bibnamefont{Xiang}}\ and\ \bibinfo
  {author} {\bibfnamefont{X.~Q.}\ \bibnamefont{Wen}},\ }%
  \bibfield{journal}{%
  \Doi{10.1088/1126-6708/2009/10/046}{\bibinfo {journal} {JHEP}}\ }%
  \textbf{\bibinfo {volume} {10}},\ \bibinfo {pages} {046} (\bibinfo {year}
  {2009}),\ \Eprint{http://arxiv.org/abs/0901.0603}{arXiv:0901.0603}%
  \bibAnnoteFile{NoStop}{Xiang:2009yq}%
\bibitem{Nozari:2012nf}%
  \BibitemOpen
  \bibfield{author}{%
  \bibinfo {author} {\bibfnamefont{K.}~\bibnamefont{Nozari}}\ and\ \bibinfo
  {author} {\bibfnamefont{S.}~\bibnamefont{Saghafi}},\ }%
  \bibfield{journal}{%
  \Doi{10.1007/JHEP11(2012)005}{\bibinfo {journal} {JHEP}}\ }%
  \textbf{\bibinfo {volume} {11}},\ \bibinfo {pages} {005} (\bibinfo {year}
  {2012}),\ \Eprint{http://arxiv.org/abs/1206.5621}{arXiv:1206.5621}%
  \bibAnnoteFile{NoStop}{Nozari:2012nf}%
\bibitem{Adler:2001vs}%
  \BibitemOpen
  \bibfield{author}{%
  \bibinfo {author} {\bibfnamefont{R.~J.}\ \bibnamefont{Adler}}, \bibinfo
  {author} {\bibfnamefont{P.}~\bibnamefont{Chen}},\ and\ \bibinfo {author}
  {\bibfnamefont{D.~I.}\ \bibnamefont{Santiago}},\ }%
  \bibfield{journal}{%
  \Doi{10.1023/A:1015281430411}{\bibinfo {journal} {Gen. Rel. Grav.}}\ }%
  \textbf{\bibinfo {volume} {33}},\ \bibinfo {pages} {2101} (\bibinfo {year}
  {2001}),\ \Eprint{http://arxiv.org/abs/gr-qc/0106080}{arXiv:gr-qc/0106080}%
  \bibAnnoteFile{NoStop}{Adler:2001vs}%
\bibitem{Zhu:2008cg}%
  \BibitemOpen
  \bibfield{author}{%
  \bibinfo {author} {\bibfnamefont{T.}~\bibnamefont{Zhu}}, \bibinfo {author}
  {\bibfnamefont{J.-R.}\ \bibnamefont{Ren}},\ and\ \bibinfo {author}
  {\bibfnamefont{M.-F.}\ \bibnamefont{Li}},\ }%
  \bibfield{journal}{%
  \Doi{10.1016/j.physletb.2009.03.020}{\bibinfo {journal} {Phys. Lett. B}}\ }%
  \textbf{\bibinfo {volume} {674}},\ \bibinfo {pages} {204} (\bibinfo {year}
  {2009}),\ \Eprint{http://arxiv.org/abs/0811.0212}{arXiv:0811.0212}%
  \bibAnnoteFile{NoStop}{Zhu:2008cg}%
\bibitem{Feng:2022gdz}%
  \BibitemOpen
  \bibfield{author}{%
  \bibinfo {author} {\bibfnamefont{Z.-W.}\ \bibnamefont{Feng}}, \bibinfo
  {author} {\bibfnamefont{X.}~\bibnamefont{Zhou}},\ and\ \bibinfo {author}
  {\bibfnamefont{S.-Q.}\ \bibnamefont{Zhou}},\ }%
  \bibfield{journal}{%
  \Doi{10.1088/1475-7516/2022/06/022}{\bibinfo {journal} {JCAP}}\ }%
  \textbf{\bibinfo {volume} {06}},\ \bibinfo {pages} {022} (\bibinfo {year}
  {2022}),\ \Eprint{http://arxiv.org/abs/2203.11671}{arXiv:2203.11671}%
  \bibAnnoteFile{NoStop}{Feng:2022gdz}%
\bibitem{Anacleto:2015mma}%
  \BibitemOpen
  \bibfield{author}{%
  \bibinfo {author} {\bibfnamefont{M.~A.}\ \bibnamefont{Anacleto}}, \bibinfo
  {author} {\bibfnamefont{F.~A.}\ \bibnamefont{Brito}},\ and\ \bibinfo {author}
  {\bibfnamefont{E.}~\bibnamefont{Passos}},\ }%
  \bibfield{journal}{%
  \Doi{10.1016/j.physletb.2015.07.072}{\bibinfo {journal} {Phys. Lett. B}}\ }%
  \textbf{\bibinfo {volume} {749}},\ \bibinfo {pages} {181} (\bibinfo {year}
  {2015}),\ \Eprint{http://arxiv.org/abs/1504.06295}{arXiv:1504.06295}%
  \bibAnnoteFile{NoStop}{Anacleto:2015mma}%
\bibitem{Awad:2014bta}%
  \BibitemOpen
  \bibfield{author}{%
  \bibinfo {author} {\bibfnamefont{A.}~\bibnamefont{Awad}}\ and\ \bibinfo
  {author} {\bibfnamefont{A.~F.}\ \bibnamefont{Ali}},\ }%
  \bibfield{journal}{%
  \Doi{10.1007/JHEP06(2014)093}{\bibinfo {journal} {JHEP}}\ }%
  \textbf{\bibinfo {volume} {06}},\ \bibinfo {pages} {093} (\bibinfo {year}
  {2014}),\ \Eprint{http://arxiv.org/abs/1404.7825}{arXiv:1404.7825}%
  \bibAnnoteFile{NoStop}{Awad:2014bta}%
\bibitem{Salah:2016kre}%
  \BibitemOpen
  \bibfield{author}{%
  \bibinfo {author} {\bibfnamefont{M.}~\bibnamefont{Salah}}, \bibinfo {author}
  {\bibfnamefont{F.}~\bibnamefont{Hammad}}, \bibinfo {author}
  {\bibfnamefont{M.}~\bibnamefont{Faizal}},\ and\ \bibinfo {author}
  {\bibfnamefont{A.~F.}\ \bibnamefont{Ali}},\ }%
  \bibfield{journal}{%
  \Doi{10.1088/1475-7516/2017/02/035}{\bibinfo {journal} {JCAP}}\ }%
  \textbf{\bibinfo {volume} {02}},\ \bibinfo {pages} {035} (\bibinfo {year}
  {2017}),\ \Eprint{http://arxiv.org/abs/1608.00560}{arXiv:1608.00560}%
  \bibAnnoteFile{NoStop}{Salah:2016kre}%
\bibitem{Khodadi:2018wed}%
  \BibitemOpen
  \bibfield{author}{%
  \bibinfo {author} {\bibfnamefont{M.}~\bibnamefont{Khodadi}}, \bibinfo
  {author} {\bibfnamefont{K.}~\bibnamefont{Nozari}},\ and\ \bibinfo {author}
  {\bibfnamefont{F.}~\bibnamefont{Hajkarim}},\ }%
  \bibfield{journal}{%
  \Doi{10.1140/epjc/s10052-018-6191-4}{\bibinfo {journal} {Eur. Phys. J. C}}\
  }%
  \textbf{\bibinfo {volume} {78}},\ \bibinfo {pages} {716} (\bibinfo {year}
  {2018}),\ \Eprint{http://arxiv.org/abs/1808.08436}{arXiv:1808.08436}%
  \bibAnnoteFile{NoStop}{Khodadi:2018wed}%
\bibitem{Khodadi:2018scn}%
  \BibitemOpen
  \bibfield{author}{%
  \bibinfo {author} {\bibfnamefont{M.}~\bibnamefont{Khodadi}}, \bibinfo
  {author} {\bibfnamefont{K.}~\bibnamefont{Nozari}}, \bibinfo {author}
  {\bibfnamefont{H.}~\bibnamefont{Abedi}},\ and\ \bibinfo {author}
  {\bibfnamefont{S.}~\bibnamefont{Capozziello}},\ }%
  \bibfield{journal}{%
  \Doi{10.1016/j.physletb.2018.07.010}{\bibinfo {journal} {Phys. Lett. B}}\ }%
  \textbf{\bibinfo {volume} {783}},\ \bibinfo {pages} {326} (\bibinfo {year}
  {2018}),\ \Eprint{http://arxiv.org/abs/1805.11310}{arXiv:1805.11310 [gr-qc]}%
  \bibAnnoteFile{NoStop}{Khodadi:2018scn}%
\bibitem{Okcu:2020ybv}%
  \BibitemOpen
  \bibfield{author}{%
  \bibinfo {author} {\bibfnamefont{O.}~\bibnamefont{\"Okc\"u}}, \bibinfo
  {author} {\bibfnamefont{C.}~\bibnamefont{Corda}},\ and\ \bibinfo {author}
  {\bibfnamefont{E.}~\bibnamefont{Aydiner}},\ }%
  \bibfield{journal}{%
  \Doi{10.1209/0295-5075/129/50002}{\bibinfo {journal} {EPL}}\ }%
  \textbf{\bibinfo {volume} {129}},\ \bibinfo {pages} {50002} (\bibinfo {year}
  {2020}),\ \Eprint{http://arxiv.org/abs/2003.11369}{arXiv:2003.11369}%
  \bibAnnoteFile{NoStop}{Okcu:2020ybv}%
\bibitem{Luo:2023rhk}%
  \BibitemOpen
  \bibfield{author}{%
  \bibinfo {author} {\bibfnamefont{S.-S.}\ \bibnamefont{Luo}}\ and\ \bibinfo
  {author} {\bibfnamefont{Z.-W.}\ \bibnamefont{Feng}},\ }%
  \bibfield{journal}{%
  \Doi{10.1016/j.aop.2023.169449}{\bibinfo {journal} {Annals Phys.}}\ }%
  \textbf{\bibinfo {volume} {458}},\ \bibinfo {pages} {169449} (\bibinfo {year}
  {2023}),\ \Eprint{http://arxiv.org/abs/2306.10078}{arXiv:2306.10078}%
  \bibAnnoteFile{NoStop}{Luo:2023rhk}%
\bibitem{Das:2021nbq}%
  \BibitemOpen
  \bibfield{author}{%
  \bibinfo {author} {\bibfnamefont{S.}~\bibnamefont{Das}}, \bibinfo {author}
  {\bibfnamefont{M.}~\bibnamefont{Fridman}}, \bibinfo {author}
  {\bibfnamefont{G.}~\bibnamefont{Lambiase}},\ and\ \bibinfo {author}
  {\bibfnamefont{E.~C.}\ \bibnamefont{Vagenas}},\ }%
  \bibfield{journal}{%
  \Doi{10.1016/j.physletb.2021.136841}{\bibinfo {journal} {Phys. Lett. B}}\ }%
  \textbf{\bibinfo {volume} {824}},\ \bibinfo {pages} {136841} (\bibinfo {year}
  {2022}),\ \Eprint{http://arxiv.org/abs/2107.02077}{arXiv:2107.02077}%
  \bibAnnoteFile{NoStop}{Das:2021nbq}%
\bibitem{Luciano:2021vkl}%
  \BibitemOpen
  \bibfield{author}{%
  \bibinfo {author} {\bibfnamefont{G.~G.}\ \bibnamefont{Luciano}},\ }%
  \bibfield{journal}{%
  \Doi{10.1140/epjc/s10052-021-09891-2}{\bibinfo {journal} {Eur. Phys. J. C}}\
  }%
  \textbf{\bibinfo {volume} {81}},\ \bibinfo {pages} {1086} (\bibinfo {year}
  {2021}),\ \Eprint{http://arxiv.org/abs/2111.06000}{arXiv:2111.06000}%
  \bibAnnoteFile{NoStop}{Luciano:2021vkl}%
\bibitem{Okcu:2024tnw}%
  \BibitemOpen
  \bibfield{author}{%
  \bibinfo {author} {\bibfnamefont{O.}~\bibnamefont{\"Okc\"u}}}%
   (\bibinfo {month} {1}\ \bibinfo {year} {2024}),\
  \Eprint{http://arxiv.org/abs/2401.09477}{arXiv:2401.09477 [gr-qc]}%
  \bibAnnoteFile{NoStop}{Okcu:2024tnw}%
\bibitem{Kolb:1981hk}%
  \BibitemOpen
  \bibfield{author}{%
  \bibinfo {author} {\bibfnamefont{E.~W.}\ \bibnamefont{Kolb}}\ and\ \bibinfo
  {author} {\bibfnamefont{M.~S.}\ \bibnamefont{Turner}},\ }%
  \bibfield{journal}{%
  \Doi{10.1038/294521a0}{\bibinfo {journal} {Nature}}\ }%
  \textbf{\bibinfo {volume} {294}},\ \bibinfo {pages} {521} (\bibinfo {year}
  {1981})%
  \bibAnnoteFile{NoStop}{Kolb:1981hk}%
\bibitem{Bernstein:1988ad}%
  \BibitemOpen
  \bibfield{author}{%
  \bibinfo {author} {\bibfnamefont{J.}~\bibnamefont{Bernstein}}, \bibinfo
  {author} {\bibfnamefont{L.~S.}\ \bibnamefont{Brown}},\ and\ \bibinfo {author}
  {\bibfnamefont{G.}~\bibnamefont{Feinberg}},\ }%
  \bibfield{journal}{%
  \Doi{10.1103/RevModPhys.61.25}{\bibinfo {journal} {Rev. Mod. Phys.}}\ }%
  \textbf{\bibinfo {volume} {61}},\ \bibinfo {pages} {25} (\bibinfo {year}
  {1989})%
  \bibAnnoteFile{NoStop}{Bernstein:1988ad}%
\bibitem{Du:2022mvr}%
  \BibitemOpen
  \bibfield{author}{%
  \bibinfo {author} {\bibfnamefont{X.-D.}\ \bibnamefont{Du}}\ and\ \bibinfo
  {author} {\bibfnamefont{C.-Y.}\ \bibnamefont{Long}},\ }%
  \bibfield{journal}{%
  \Doi{10.1007/JHEP10(2022)063}{\bibinfo {journal} {JHEP}}\ }%
  \textbf{\bibinfo {volume} {10}},\ \bibinfo {pages} {063} (\bibinfo {year}
  {2022}),\ \Eprint{http://arxiv.org/abs/2208.12918}{arXiv:2208.12918}%
  \bibAnnoteFile{NoStop}{Du:2022mvr}%
\bibitem{Cai:2008ys}%
  \BibitemOpen
  \bibfield{author}{%
  \bibinfo {author} {\bibfnamefont{R.-G.}\ \bibnamefont{Cai}}, \bibinfo
  {author} {\bibfnamefont{L.-M.}\ \bibnamefont{Cao}},\ and\ \bibinfo {author}
  {\bibfnamefont{Y.-P.}\ \bibnamefont{Hu}},\ }%
  \bibfield{journal}{%
  \Doi{10.1088/1126-6708/2008/08/090}{\bibinfo {journal} {JHEP}}\ }%
  \textbf{\bibinfo {volume} {08}},\ \bibinfo {pages} {090} (\bibinfo {year}
  {2008}),\ \Eprint{http://arxiv.org/abs/0807.1232}{arXiv:0807.1232}%
  \bibAnnoteFile{NoStop}{Cai:2008ys}%
\bibitem{Bargueno:2015tea}%
  \BibitemOpen
  \bibfield{author}{%
  \bibinfo {author} {\bibfnamefont{P.}~\bibnamefont{Bargue\~no}}\ and\ \bibinfo
  {author} {\bibfnamefont{E.~C.}\ \bibnamefont{Vagenas}},\ }%
  \bibfield{journal}{%
  \Doi{10.1016/j.physletb.2015.01.016}{\bibinfo {journal} {Phys. Lett. B}}\ }%
  \textbf{\bibinfo {volume} {742}},\ \bibinfo {pages} {15} (\bibinfo {year}
  {2015}),\ \Eprint{http://arxiv.org/abs/1501.03256}{arXiv:1501.03256}%
  \bibAnnoteFile{NoStop}{Bargueno:2015tea}%
\bibitem{Sheykhi:2010wm}%
  \BibitemOpen
  \bibfield{author}{%
  \bibinfo {author} {\bibfnamefont{A.}~\bibnamefont{Sheykhi}},\ }%
  \bibfield{journal}{%
  \Doi{10.1103/PhysRevD.81.104011}{\bibinfo {journal} {Phys. Rev. D}}\ }%
  \textbf{\bibinfo {volume} {81}},\ \bibinfo {pages} {104011} (\bibinfo {year}
  {2010}),\ \Eprint{http://arxiv.org/abs/1004.0627}{arXiv:1004.0627}%
  \bibAnnoteFile{NoStop}{Sheykhi:2010wm}%
\bibitem{Feng:2017vvw}%
  \BibitemOpen
  \bibfield{author}{%
  \bibinfo {author} {\bibfnamefont{Z.-W.}\ \bibnamefont{Feng}}\ and\ \bibinfo
  {author} {\bibfnamefont{S.-Z.}\ \bibnamefont{Yang}},\ }%
  \bibfield{journal}{%
  \Doi{10.1155/2018/5968284}{\bibinfo {journal} {Adv. High Energy Phys.}}\ }%
  \textbf{\bibinfo {volume} {2018}},\ \bibinfo {pages} {5968284} (\bibinfo
  {year} {2018}),\ \Eprint{http://arxiv.org/abs/1708.08324}{arXiv:1708.08324}%
  \bibAnnoteFile{NoStop}{Feng:2017vvw}%
\bibitem{Suh:1999va}%
  \BibitemOpen
  \bibfield{author}{%
  \bibinfo {author} {\bibfnamefont{I.-S.}\ \bibnamefont{Suh}}\ and\ \bibinfo
  {author} {\bibfnamefont{G.~J.}\ \bibnamefont{Mathews}},\ }%
  \bibfield{journal}{%
  \Doi{10.1103/PhysRevD.59.123002}{\bibinfo {journal} {Phys. Rev. D}}\ }%
  \textbf{\bibinfo {volume} {59}},\ \bibinfo {pages} {123002} (\bibinfo {year}
  {1999}),\
  \Eprint{http://arxiv.org/abs/astro-ph/9812185}{arXiv:astro-ph/9812185}%
  \bibAnnoteFile{NoStop}{Suh:1999va}%
\bibitem{Ghoshal:2021ief}%
  \BibitemOpen
  \bibfield{author}{%
  \bibinfo {author} {\bibfnamefont{A.}~\bibnamefont{Ghoshal}}\ and\ \bibinfo
  {author} {\bibfnamefont{G.}~\bibnamefont{Lambiase}}}%
   (\bibinfo {month} {4}\ \bibinfo {year} {2021}),\
  \Eprint{http://arxiv.org/abs/2104.11296}{arXiv:2104.11296}%
  \bibAnnoteFile{NoStop}{Ghoshal:2021ief}%
\bibitem{Capozziello:2017bxm}%
  \BibitemOpen
  \bibfield{author}{%
  \bibinfo {author} {\bibfnamefont{S.}~\bibnamefont{Capozziello}}, \bibinfo
  {author} {\bibfnamefont{G.}~\bibnamefont{Lambiase}},\ and\ \bibinfo {author}
  {\bibfnamefont{E.~N.}\ \bibnamefont{Saridakis}},\ }%
  \bibfield{journal}{%
  \Doi{10.1140/epjc/s10052-017-5143-8}{\bibinfo {journal} {Eur. Phys. J. C}}\
  }%
  \textbf{\bibinfo {volume} {77}},\ \bibinfo {pages} {576} (\bibinfo {year}
  {2017}),\ \Eprint{http://arxiv.org/abs/1702.07952}{arXiv:1702.07952}%
  \bibAnnoteFile{NoStop}{Capozziello:2017bxm}%
\bibitem{Lambiase:2012fv}%
  \BibitemOpen
  \bibfield{author}{%
  \bibinfo {author} {\bibfnamefont{G.}~\bibnamefont{Lambiase}},\ }%
  \bibfield{journal}{%
  \Doi{10.1088/1475-7516/2012/10/028}{\bibinfo {journal} {JCAP}}\ }%
  \textbf{\bibinfo {volume} {10}},\ \bibinfo {pages} {028} (\bibinfo {year}
  {2012}),\ \Eprint{http://arxiv.org/abs/1208.5512}{arXiv:1208.5512}%
  \bibAnnoteFile{NoStop}{Lambiase:2012fv}%
\bibitem{Fields:2011zzb}%
  \BibitemOpen
  \bibfield{author}{%
  \bibinfo {author} {\bibfnamefont{B.~D.}\ \bibnamefont{Fields}},\ }%
  \bibfield{journal}{%
  \Doi{10.1146/annurev-nucl-102010-130445}{\bibinfo {journal} {Ann. Rev. Nucl.
  Part. Sci.}}\ }%
  \textbf{\bibinfo {volume} {61}},\ \bibinfo {pages} {47} (\bibinfo {year}
  {2011}),\ \Eprint{http://arxiv.org/abs/1203.3551}{arXiv:1203.3551}%
  \bibAnnoteFile{NoStop}{Fields:2011zzb}%
\bibitem{Boran:2013pux}%
  \BibitemOpen
  \bibfield{author}{%
  \bibinfo {author} {\bibfnamefont{S.}~\bibnamefont{Boran}}\ and\ \bibinfo
  {author} {\bibfnamefont{E.~O.}\ \bibnamefont{Kahya}},\ }%
  \bibfield{journal}{%
  \Doi{10.1155/2014/282675}{\bibinfo {journal} {Adv. High Energy Phys.}}\ }%
  \textbf{\bibinfo {volume} {2014}},\ \bibinfo {pages} {282675} (\bibinfo
  {year} {2014}),\ \Eprint{http://arxiv.org/abs/1310.6145}{arXiv:1310.6145}%
  \bibAnnoteFile{NoStop}{Boran:2013pux}%
\bibitem{Kneller:2004jz}%
  \BibitemOpen
  \bibfield{author}{%
  \bibinfo {author} {\bibfnamefont{J.~P.}\ \bibnamefont{Kneller}}\ and\
  \bibinfo {author} {\bibfnamefont{G.}~\bibnamefont{Steigman}},\ }%
  \bibfield{journal}{%
  \Doi{10.1088/1367-2630/6/1/117}{\bibinfo {journal} {New J. Phys.}}\ }%
  \textbf{\bibinfo {volume} {6}},\ \bibinfo {pages} {117} (\bibinfo {year}
  {2004}),\
  \Eprint{http://arxiv.org/abs/astro-ph/0406320}{arXiv:astro-ph/0406320}%
  \bibAnnoteFile{NoStop}{Kneller:2004jz}%
\bibitem{Steigman:2007xt}%
  \BibitemOpen
  \bibfield{author}{%
  \bibinfo {author} {\bibfnamefont{G.}~\bibnamefont{Steigman}},\ }%
  \bibfield{journal}{%
  \Doi{10.1146/annurev.nucl.56.080805.140437}{\bibinfo {journal} {Ann. Rev.
  Nucl. Part. Sci.}}\ }%
  \textbf{\bibinfo {volume} {57}},\ \bibinfo {pages} {463} (\bibinfo {year}
  {2007}),\ \Eprint{http://arxiv.org/abs/0712.1100}{arXiv:0712.1100}%
  \bibAnnoteFile{NoStop}{Steigman:2007xt}%
\bibitem{Steigman:2012ve}%
  \BibitemOpen
  \bibfield{author}{%
  \bibinfo {author} {\bibfnamefont{G.}~\bibnamefont{Steigman}},\ }%
  \bibfield{journal}{%
  \Doi{10.1155/2012/268321}{\bibinfo {journal} {Adv. High Energy Phys.}}\ }%
  \textbf{\bibinfo {volume} {2012}},\ \bibinfo {pages} {268321} (\bibinfo
  {year} {2012}),\ \Eprint{http://arxiv.org/abs/1208.0032}{arXiv:1208.0032}%
  \bibAnnoteFile{NoStop}{Steigman:2012ve}%
\bibitem{Simha:2008zj}%
  \BibitemOpen
  \bibfield{author}{%
  \bibinfo {author} {\bibfnamefont{V.}~\bibnamefont{Simha}}\ and\ \bibinfo
  {author} {\bibfnamefont{G.}~\bibnamefont{Steigman}},\ }%
  \bibfield{journal}{%
  \Doi{10.1088/1475-7516/2008/06/016}{\bibinfo {journal} {JCAP}}\ }%
  \textbf{\bibinfo {volume} {06}},\ \bibinfo {pages} {016} (\bibinfo {year}
  {2008}),\ \Eprint{http://arxiv.org/abs/0803.3465}{arXiv:0803.3465}%
  \bibAnnoteFile{NoStop}{Simha:2008zj}%
\bibitem{Schoneberg:2024ifp}%
  \BibitemOpen
  \bibfield{author}{%
  \bibinfo {author} {\bibfnamefont{N.}~\bibnamefont{Sch\"oneberg}}}%
   (\bibinfo {month} {1}\ \bibinfo {year} {2024}),\
  \Eprint{http://arxiv.org/abs/2401.15054}{arXiv:2401.15054 [astro-ph.CO]}%
  \bibAnnoteFile{NoStop}{Schoneberg:2024ifp}%
\bibitem{WMAP:2010qai}%
  \BibitemOpen
  \bibfield{author}{%
  \bibinfo {author} {\bibfnamefont{E.}~\bibnamefont{Komatsu}} \emph{et~al.}
  (\bibinfo {collaboration} {WMAP}),\ }%
  \bibfield{journal}{%
  \Doi{10.1088/0067-0049/192/2/18}{\bibinfo {journal} {Astrophys. J. Suppl.}}\
  }%
  \textbf{\bibinfo {volume} {192}},\ \bibinfo {pages} {18} (\bibinfo {year}
  {2011}),\ \Eprint{http://arxiv.org/abs/1001.4538}{arXiv:1001.4538
  [astro-ph.CO]}%
  \bibAnnoteFile{NoStop}{WMAP:2010qai}%
\bibitem{Fields:2019pfx}%
  \BibitemOpen
  \bibfield{author}{%
  \bibinfo {author} {\bibfnamefont{B.~D.}\ \bibnamefont{Fields}}, \bibinfo
  {author} {\bibfnamefont{K.~A.}\ \bibnamefont{Olive}}, \bibinfo {author}
  {\bibfnamefont{T.-H.}\ \bibnamefont{Yeh}},\ and\ \bibinfo {author}
  {\bibfnamefont{C.}~\bibnamefont{Young}},\ }%
  \bibfield{journal}{%
  \Doi{10.1088/1475-7516/2020/03/010}{\bibinfo {journal} {JCAP}}\ }%
  \textbf{\bibinfo {volume} {03}},\ \bibinfo {pages} {010} (\bibinfo {year}
  {2020}),\ \bibinfo {note} {[Erratum: JCAP 11, E02 (2020)]},\
  \Eprint{http://arxiv.org/abs/1912.01132}{arXiv:1912.01132}%
  \bibAnnoteFile{NoStop}{Fields:2019pfx}%
\bibitem{Bhattacharjee:2020uhs}%
  \BibitemOpen
  \bibfield{author}{%
  \bibinfo {author} {\bibfnamefont{S.}~\bibnamefont{Bhattacharjee}}\ and\
  \bibinfo {author} {\bibfnamefont{P.~K.}\ \bibnamefont{Sahoo}},\ }%
  \bibfield{journal}{%
  \Doi{10.1140/epjp/s13360-020-00361-4}{\bibinfo {journal} {Eur. Phys. J.
  Plus}}\ }%
  \textbf{\bibinfo {volume} {135}},\ \bibinfo {pages} {350} (\bibinfo {year}
  {2020}),\ \Eprint{http://arxiv.org/abs/2004.04684}{arXiv:2004.04684}%
  \bibAnnoteFile{NoStop}{Bhattacharjee:2020uhs}%
\bibitem{Kouwn:2018rmp}%
  \BibitemOpen
  \bibfield{author}{%
  \bibinfo {author} {\bibfnamefont{S.}~\bibnamefont{Kouwn}},\ }%
  \bibfield{journal}{%
  \Doi{10.1016/j.dark.2018.07.001}{\bibinfo {journal} {Phys. Dark Univ.}}\ }%
  \textbf{\bibinfo {volume} {21}},\ \bibinfo {pages} {76} (\bibinfo {year}
  {2018}),\ \Eprint{http://arxiv.org/abs/1805.07278}{arXiv:1805.07278
  [astro-ph.CO]}%
  \bibAnnoteFile{NoStop}{Kouwn:2018rmp}%
\bibitem{Giardino:2020myz}%
  \BibitemOpen
  \bibfield{author}{%
  \bibinfo {author} {\bibfnamefont{S.}~\bibnamefont{Giardino}}\ and\ \bibinfo
  {author} {\bibfnamefont{V.}~\bibnamefont{Salzano}},\ }%
  \bibfield{journal}{%
  \Doi{10.1140/epjc/s10052-021-08914-2}{\bibinfo {journal} {Eur. Phys. J. C}}\
  }%
  \textbf{\bibinfo {volume} {81}},\ \bibinfo {pages} {110} (\bibinfo {year}
  {2021}),\ \Eprint{http://arxiv.org/abs/2006.01580}{arXiv:2006.01580}%
  \bibAnnoteFile{NoStop}{Giardino:2020myz}%
\bibitem{Luciano:2022knb}%
  \BibitemOpen
  \bibfield{author}{%
  \bibinfo {author} {\bibfnamefont{G.~G.}\ \bibnamefont{Luciano}},\ }%
  \bibfield{journal}{%
  \Doi{10.1140/epjc/s10052-022-10285-1}{\bibinfo {journal} {Eur. Phys. J. C}}\
  }%
  \textbf{\bibinfo {volume} {82}},\ \bibinfo {pages} {314} (\bibinfo {year}
  {2022})%
  \bibAnnoteFile{NoStop}{Luciano:2022knb}%
\bibitem{Gine:2020izd}%
  \BibitemOpen
  \bibfield{author}{%
  \bibinfo {author} {\bibfnamefont{J.}~\bibnamefont{Gin\'e}}\ and\ \bibinfo
  {author} {\bibfnamefont{G.~G.}\ \bibnamefont{Luciano}},\ }%
  \bibfield{journal}{%
  \Doi{10.1140/epjc/s10052-020-08636-x}{\bibinfo {journal} {Eur. Phys. J. C}}\
  }%
  \textbf{\bibinfo {volume} {80}},\ \bibinfo {pages} {1039} (\bibinfo {year}
  {2020})%
  \bibAnnoteFile{NoStop}{Gine:2020izd}%
\bibitem{Ali:2011fa}%
  \BibitemOpen
  \bibfield{author}{%
  \bibinfo {author} {\bibfnamefont{A.~F.}\ \bibnamefont{Ali}}, \bibinfo
  {author} {\bibfnamefont{S.}~\bibnamefont{Das}},\ and\ \bibinfo {author}
  {\bibfnamefont{E.~C.}\ \bibnamefont{Vagenas}},\ }%
  \bibfield{journal}{%
  \Doi{10.1103/PhysRevD.84.044013}{\bibinfo {journal} {Phys. Rev. D}}\ }%
  \textbf{\bibinfo {volume} {84}},\ \bibinfo {pages} {044013} (\bibinfo {year}
  {2011}),\ \Eprint{http://arxiv.org/abs/1107.3164}{arXiv:1107.3164 [hep-th]}%
  \bibAnnoteFile{NoStop}{Ali:2011fa}%
\bibitem{Ghosh:2013qra}%
  \BibitemOpen
  \bibfield{author}{%
  \bibinfo {author} {\bibfnamefont{S.}~\bibnamefont{Ghosh}},\ }%
  \bibfield{journal}{%
  \Doi{10.1088/0264-9381/31/2/025025}{\bibinfo {journal} {Class. Quant.
  Grav.}}\ }%
  \textbf{\bibinfo {volume} {31}},\ \bibinfo {pages} {025025} (\bibinfo {year}
  {2014}),\ \Eprint{http://arxiv.org/abs/1303.1256}{arXiv:1303.1256 [gr-qc]}%
  \bibAnnoteFile{NoStop}{Ghosh:2013qra}%
\bibitem{Bushev:2019zvw}%
  \BibitemOpen
  \bibfield{author}{%
  \bibinfo {author} {\bibfnamefont{P.~A.}\ \bibnamefont{Bushev}}, \bibinfo
  {author} {\bibfnamefont{J.}~\bibnamefont{Bourhill}}, \bibinfo {author}
  {\bibfnamefont{M.}~\bibnamefont{Goryachev}}, \bibinfo {author}
  {\bibfnamefont{N.}~\bibnamefont{Kukharchyk}}, \bibinfo {author}
  {\bibfnamefont{E.}~\bibnamefont{Ivanov}}, \bibinfo {author}
  {\bibfnamefont{S.}~\bibnamefont{Galliou}}, \bibinfo {author}
  {\bibfnamefont{M.~E.}\ \bibnamefont{Tobar}},\ and\ \bibinfo {author}
  {\bibfnamefont{S.}~\bibnamefont{Danilishin}},\ }%
  \bibfield{journal}{%
  \Doi{10.1103/PhysRevD.100.066020}{\bibinfo {journal} {Phys. Rev. D}}\ }%
  \textbf{\bibinfo {volume} {100}},\ \bibinfo {pages} {066020} (\bibinfo {year}
  {2019}),\ \Eprint{http://arxiv.org/abs/1903.03346}{arXiv:1903.03346
  [quant-ph]}%
  \bibAnnoteFile{NoStop}{Bushev:2019zvw}%
\bibitem{Bambi:2007ty}%
  \BibitemOpen
  \bibfield{author}{%
  \bibinfo {author} {\bibfnamefont{C.}~\bibnamefont{Bambi}}\ and\ \bibinfo
  {author} {\bibfnamefont{F.~R.}\ \bibnamefont{Urban}},\ }%
  \bibfield{journal}{%
  \Doi{10.1088/0264-9381/25/9/095006}{\bibinfo {journal} {Class. Quant.
  Grav.}}\ }%
  \textbf{\bibinfo {volume} {25}},\ \bibinfo {pages} {095006} (\bibinfo {year}
  {2008}),\ \Eprint{http://arxiv.org/abs/0709.1965}{arXiv:0709.1965}%
  \bibAnnoteFile{NoStop}{Bambi:2007ty}%
\bibitem{Kempf:1994su}%
  \BibitemOpen
  \bibfield{author}{%
  \bibinfo {author} {\bibfnamefont{A.}~\bibnamefont{Kempf}}, \bibinfo {author}
  {\bibfnamefont{G.}~\bibnamefont{Mangano}},\ and\ \bibinfo {author}
  {\bibfnamefont{R.~B.}\ \bibnamefont{Mann}},\ }%
  \bibfield{journal}{%
  \Doi{10.1103/PhysRevD.52.1108}{\bibinfo {journal} {Phys. Rev. D}}\ }%
  \textbf{\bibinfo {volume} {52}},\ \bibinfo {pages} {1108} (\bibinfo {year}
  {1995}),\ \Eprint{http://arxiv.org/abs/hep-th/9412167}{arXiv:hep-th/9412167}%
  \bibAnnoteFile{NoStop}{Kempf:1994su}%
\bibitem{Mureika:2018gxl}%
  \BibitemOpen
  \bibfield{author}{%
  \bibinfo {author} {\bibfnamefont{J.~R.}\ \bibnamefont{Mureika}},\ }%
  \bibfield{journal}{%
  \Doi{10.1016/j.physletb.2018.12.009}{\bibinfo {journal} {Phys. Lett. B}}\ }%
  \textbf{\bibinfo {volume} {789}},\ \bibinfo {pages} {88} (\bibinfo {year}
  {2019}),\ \Eprint{http://arxiv.org/abs/1812.01999}{arXiv:1812.01999}%
  \bibAnnoteFile{NoStop}{Mureika:2018gxl}%
\bibitem{Okcu:2022sio}%
  \BibitemOpen
  \bibfield{author}{%
  \bibinfo {author} {\bibfnamefont{O.}~\bibnamefont{\"Okc\"u}}\ and\ \bibinfo
  {author} {\bibfnamefont{E.}~\bibnamefont{Aydiner}},\ }%
  \bibfield{journal}{%
  \Doi{10.1209/0295-5075/ac6976}{\bibinfo {journal} {EPL}}\ }%
  \textbf{\bibinfo {volume} {138}},\ \bibinfo {pages} {39002} (\bibinfo {year}
  {2022}),\ \Eprint{http://arxiv.org/abs/2209.03170}{arXiv:2209.03170}%
  \bibAnnoteFile{NoStop}{Okcu:2022sio}%
\bibitem{Moradpour:2019yiq}%
  \BibitemOpen
  \bibfield{author}{%
  \bibinfo {author} {\bibfnamefont{H.}~\bibnamefont{Moradpour}}, \bibinfo
  {author} {\bibfnamefont{C.}~\bibnamefont{Corda}}, \bibinfo {author}
  {\bibfnamefont{A.~H.}\ \bibnamefont{Ziaie}},\ and\ \bibinfo {author}
  {\bibfnamefont{S.}~\bibnamefont{Ghaffari}},\ }%
  \bibfield{journal}{%
  \Doi{10.1209/0295-5075/127/60006}{\bibinfo {journal} {EPL}}\ }%
  \textbf{\bibinfo {volume} {127}},\ \bibinfo {pages} {60006} (\bibinfo {year}
  {2019}),\ \Eprint{http://arxiv.org/abs/1902.01703}{arXiv:1902.01703}%
  \bibAnnoteFile{NoStop}{Moradpour:2019yiq}%
\bibitem{Chung:2019iwp}%
  \BibitemOpen
  \bibfield{author}{%
  \bibinfo {author} {\bibfnamefont{W.~S.}\ \bibnamefont{Chung}}\ and\ \bibinfo
  {author} {\bibfnamefont{H.}~\bibnamefont{Hassanabadi}},\ }%
  \bibfield{journal}{%
  \Doi{10.1016/j.physletb.2019.04.063}{\bibinfo {journal} {Phys. Lett. B}}\ }%
  \textbf{\bibinfo {volume} {793}},\ \bibinfo {pages} {451} (\bibinfo {year}
  {2019})%
  \bibAnnoteFile{NoStop}{Chung:2019iwp}%
\bibitem{Dabrowski:2019wjk}%
  \BibitemOpen
  \bibfield{author}{%
  \bibinfo {author} {\bibfnamefont{M.~P.}\ \bibnamefont{Dabrowski}}\ and\
  \bibinfo {author} {\bibfnamefont{F.}~\bibnamefont{Wagner}},\ }%
  \bibfield{journal}{%
  \Doi{10.1140/epjc/s10052-019-7232-3}{\bibinfo {journal} {Eur. Phys. J. C}}\
  }%
  \textbf{\bibinfo {volume} {79}},\ \bibinfo {pages} {716} (\bibinfo {year}
  {2019}),\ \Eprint{http://arxiv.org/abs/1905.09713}{arXiv:1905.09713}%
  \bibAnnoteFile{NoStop}{Dabrowski:2019wjk}%
\end{thebibliography}%

\end{document}